\documentclass[10pt, conference, letterpaper]{ IEEEtran}
\IEEEoverridecommandlockouts

\usepackage{cite}
\usepackage{amsmath,amssymb,amsfonts}
\usepackage{algorithm}
\usepackage{algorithmic}
\usepackage{graphicx}
\usepackage{textcomp}
\usepackage{xcolor}
\usepackage{booktabs}
\usepackage{multirow}
\usepackage{threeparttable}
\usepackage[table]{xcolor}

\def\BibTeX{{\rm B\kern-.05em{\sc i\kern-.025em b}\kern-.08em
    T\kern-.1667em\lower.7ex\hbox{E}\kern-.125emX}}
\begin{document}

\title{RadioDreamer: A World Model for Radio Map Forecasting and Imagination-Based Optimization\\


\author{\IEEEauthorblockN{Haoye Chai, Mugen Peng}}
}

\maketitle

\begin{abstract}
Radio maps characterize the spatial distribution of wireless conditions and are
fundamental to network planning and optimization. However, existing learning-based
methods reconstruct isolated radio-map snapshots and do not model how network
actions drive the evolution of the radio environment.
In this paper, we present RadioDreamer, a two-stage world model that unifies action-conditioned radio-map forecasting and imagination-based network optimization within a latent space. To learn causal environmental transitions from sequential interactions, we introduce a structured spatio-temporal transformer backbone coupled with dynamic-region supervision, enabling accurate modeling of action-sensitive areas. Second, rather than relying on conventional physical trial-and-error, RadioDreamer establishes a general planning paradigm that generates counterfactual latent trajectories to evaluate and optimize long-horizon network parameter configurations. 
Experiments show that RadioDreamer improves both multi-step rollout
forecasting and closed-loop network optimization. It reduces full-map
forecasting error by 25.2\% on average and achieves approximately 80\%
accuracy in predicting the evolution direction of radio signals. More
importantly, for long-horizon network configuration, imagination-based
planning surpasses model-free policies optimized directly on radio-map
observations by 8.8\% in accumulated return.
\end{abstract}


\section{Introduction}
\label{sec:introduction}

The radio propagation environment determines how wireless signals propagate through physical spaces and is essential for reliable mobile network connectivity. Radio map (RM), by representing the spatial distribution of radio conditions, has become a key enabler for network planning, beam configuration, and radio resource management~\cite{romero2022radio,zeng2021ckm,jaensch2026radiomap}.
Constructing them accurately with low cost has
received considerable attention, and learning-based methods have become the prevailing approach.
For example, RadioUNet~\cite{levie2021radiounet} and PMNet~\cite{lee2023pmnet} adopt
encoder--decoder networks to regress path-loss distributions from building
layouts and transmitter information, and RadioDiff~\cite{wang2024radiodiff}
utilizes conditional generation to
recover fine-grained radio maps.
Despite their different architectures, these methods typically model the
relationship among a static urban scene, a fixed set of network parameters,
and the corresponding radio state, learning statistical mappings
from a given scene and network configuration to a radio-map
snapshot.

\emph{ This raises a fundamental question: when the wireless environment continuously evolves in response to network actions, are isolated snapshot predictions sufficient to understand network evolution and support configuration optimization?}


The fundamental limitation of this static paradigm is that it cannot capture
the causal relationship between network actions and radio-environment changes.
The radio environment is jointly determined by the propagation scene and
network configurations: adjusting transmit power or antenna tilt reshapes
signal strength, coverage boundaries, and interference conditions~\cite{tekgul2024antenna}. By treating
different configurations as independent samples, static models learn only
statistical correlations and cannot identify changes directly induced by
network actions. Meanwhile, moving vehicles/pedestrians continuously alter
blockage, reflection, and propagation paths, creating dynamics
across spatial regions~\cite{boban2019vehicle,maccartney2017human}. Such evolution cannot be adequately represented by
isolated snapshots, limiting multi-step rollout accuracy. More importantly,
tasks such as antenna tuning, coverage optimization, and energy-saving
control require counterfactual rollouts of candidate actions and comparisons of
their long-term returns~\cite{tiwari2026digitaltwin,li2020modelbased,park2022mbrl}. Without the ability to predict radio-environment
evolution, a model may produce inaccurate long-horizon estimates and
suboptimal policies.

Recent world models, such as Dreamer~\cite{hafner2020dreamer} and V-JEPA~\cite{bardes2024revisiting}, have emerged as a paradigm for learning environment dynamics. Unlike models that fit static input--output mappings, world models learn action-conditioned state dynamics from sequential data. Their latent transitions enable recursive rollout under candidate actions, generating imagined trajectories for evaluating long-term effects. This paradigm has been applied to embodied intelligence~\cite{wu2023daydreamer,zhou2024robodreamer}, robotic control~\cite{hansen2024tdmpc2}, and autonomous driving~\cite{hu2022mile}, reducing real-world trial-and-error and improving long-horizon decision-making. More importantly, it unifies prediction, counterfactual analysis, and planning in a shared latent space, allowing candidate actions to be efficiently simulated, compared, and optimized before execution.

Motivated by this paradigm, we propose \textbf{RadioDreamer}, a world model for radio map rollout and network parameter optimization. It comprises two stages: dynamics learning and imagination-based planning. In the first stage, RadioDreamer learns latent representations and action-conditioned transitions from sequential radio-environment observations and network control actions. A \textbf{Cellular-Aware Transformer Block} captures temporal causal dependencies, inter-cell interactions, and spatial propagation patterns. We introduce \textbf{Region-Adaptive Supervision}, which adjusts supervision according to local dynamics and emphasizes regions sensitive to network actions. In the second stage, RadioDreamer recursively rolls out imagined trajectories for candidate actions and optimizes network configurations based on their long-term performance. It therefore not only forecasts how network actions reshape the radio environment, but also supports long-horizon parameter optimization through learned latent dynamics.
The main contributions of this work are summarized as follows:

\begin{itemize}
    \item To the best of our knowledge, RadioDreamer is the first world model
    for action-conditioned radio-map forecasting and network optimization in
    cellular networks. It unifies latent dynamics learning and
    imagination-based optimization: the former models multi-step radio-map
    evolution under BS actions and environmental changes, while the latter
    rolls out latent trajectories to optimize long-term network configurations.

    \item We design a Cellular-Aware Transformer Block that structurally models
    temporal causal dependencies, inter-BS interactions, and spatial
    propagation. We further introduce Region-Adaptive Supervision to emphasize
    dynamic and action-sensitive regions, improving localized dynamics
    prediction and reducing error accumulation during multi-step rollout.

    \item We evaluate RadioDreamer across three urban scenes and three
    generalization protocols. It reduces full-map forecasting error by 25.2\%
    on average and dynamic-region error by up to 45.2\%. Imagination-based
    planning further improves the episode return by 8.8\% over RM-based optimization.
\end{itemize}

\section{Related Work}
\label{sec:related_work}

Radio-map prediction has been extensively studied. Early approaches relied on
repeated measurements or propagation models, but their data-collection and
computational costs limit continuous forecasting
~\cite{gao2023kriging,romero2022radio}. Recent learning-based methods
have directly mapped environmental conditions to radio maps. RadioUNet
~\cite{levie2021radiounet} and PMNet~\cite{lee2023pmnet} predict path-loss
distributions from building layouts and transmitter information, while
RadioDiff~\cite{wang2024radiodiff} formulates radio-map construction as
conditional generation. Subsequent studies further incorporate electromagnetic
knowledge for multipath modeling~\cite{wang2025radiodiffk2}, and foundation
models exploit large-scale self-supervised pre-training to learn 
radio-environment representations~\cite{yang2026fmrme,alikhani2025lwm, 11298194}.

World models learn compact environment representations and action-conditioned
transition dynamics from sequential observations, enabling future states to be
imagined under alternative actions. Dreamer trains control policies from
latent imagined trajectories~\cite{hafner2025dreamerv3}, while V-JEPA~2
supports action-conditioned robotic planning~\cite{assran2025vjepa2}. In
autonomous driving, GAIA-1~\cite{hu2023gaia1}, Drive-WM
~\cite{wang2024drivewm}, and DrivingWorld~\cite{hu2024drivingworld} generate
controllable future scenes for trajectory planning. These works
demonstrate the capability of world models to unify prediction,
counterfactual rollout, and action planning.

Despite these advances, existing radio-map methods mainly learn static mappings
from scenes, transmitter configurations, or measurements to individual radio
maps. They reconstruct radio conditions under fixed settings but do not
explicitly model their evolution under network actions and environmental
changes. Network optimization, however, requires rolling out candidate actions
and comparing their long-term consequences. To address this gap, we propose
RadioDreamer, an action-conditioned world model for dynamic radio-map
forecasting and counterfactual network optimization.

\section{Problem Formulation}
\label{sec:problem_formulation}

\begin{figure}[ht]
\centerline{\includegraphics[width=.8\linewidth]{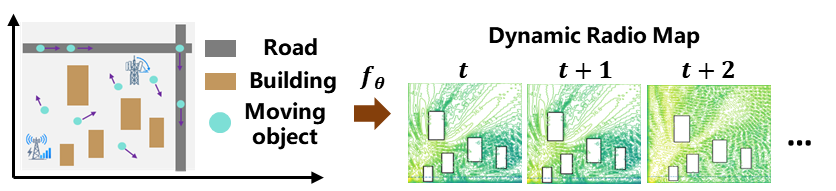}}
\caption{System model of radio map forecasting.}
\label{problem_for}
\vspace{-0.2cm}
\end{figure}

As illustrated in Fig.~\ref{problem_for}, we consider an urban wireless
environment discretized into an $R\times C$ spatial grid. The environment
contains static buildings, $B$ base stations (BSs), and moving objects such as
vehicles and pedestrians. The static scene is denoted by $\mathbf{S}$ and
includes building geometry, material properties, and BS locations. At time
slot $t$, the moving-object states $\mathbf{N}_t$ describe their locations and
motion information, whose variations continuously affect blockage, reflection,
and propagation paths. The action of the $b$-th BS is defined as
\begin{equation}
    \mathbf{a}_{t}^{b}
    =
    \left[
        p_{t}^{b},
        \delta_{t}^{b}
    \right],
    \qquad b=1,\ldots,B,
\end{equation}
where $p_{t}^{b}$ and $\delta_{t}^{b}$ denote the transmit power and antenna
downtilt, respectively.
Under the joint effects of $\mathbf{S}$, $\mathbf{N}_t$, and the BS actions,
the radio map at time slot $t$ is represented by
\begin{equation}
    \mathbf{M}_{t}
    \in \mathbb{R}^{R\times C},
\end{equation}
where $\mathbf{M}_t$ characterizes the spatial distributions of reference
signal received power (RSRP) and signal-to-interference-plus-noise ratio
(SINR). BS reconfiguration alters serving-signal strength, coverage boundaries,
and inter-cell interference, while moving objects introduce additional
time-varying propagation effects. Consequently, the radio-map sequence
$\{\mathbf{M}_t\}$ evolves according to the coupled dynamics of network actions
and environmental mobility.

Given the historical radio maps $\mathbf{M}_{t-T:t}$, BS actions
$\mathbf{a}_{t-T:t}$, moving-object states $\mathbf{N}_{t-T:t}$, and static
scene $\mathbf{S}$, we aim to learn a world model $f_\theta$ for
radio-map evolution. The historical context is defined as
\begin{equation}
    \mathcal{H}_{t}
    =
    \left\{
        \mathbf{S},
        \mathbf{M}_{t-T:t},
        \mathbf{a}_{t-T:t},
        \mathbf{N}_{t-T:t}
    \right\}.
\end{equation}
Given a future BS action $\mathbf{a}_{t+1:t+T'}$, the model generates
the corresponding future radio maps $\mathbf{M}_{t+1:t+T'}$ by learning
\begin{equation}
\begin{aligned}
    f_{\theta}
    \left(
        \mathcal{H}_{t},
        \mathbf{a}_{t+1:t+T'}
    \right)
    =
    p_{\theta}
    \left(
        \mathbf{M}_{t+1:t+T'}
        \mid
        \mathcal{H}_{t},
        \mathbf{a}_{t+1:t+T'}
    \right),
\end{aligned}
\label{eq:world_model}
\end{equation}
where $\theta$ denotes the model parameters and $T'$ is the rollout horizon.
Future moving object states are unavailable and  inferred from the historical sequence $\mathbf{N}_{t-T:t}$.


Learning such a world model is challenging because radio-map evolution depends on heterogeneous interactions among static buildings, BSs, moving objects, and network actions across spatial, temporal, and BS dimensions. BS actions can cause large changes near coverage boundaries and interference-sensitive regions, while vehicles and pedestrians continuously alter local blockage, reflection, and propagation paths. Inadequate modeling of these dynamics can lead to error accumulation during recursive rollout and degrade long-horizon accuracy.

\section{The Framework of RadioDreamer}
\label{sec:radiodreamer}

\subsection{RadioDreamer Paradigm}
\label{subsec:paradigm}

\begin{figure}[t]
\centerline{\includegraphics[width=.8\linewidth]{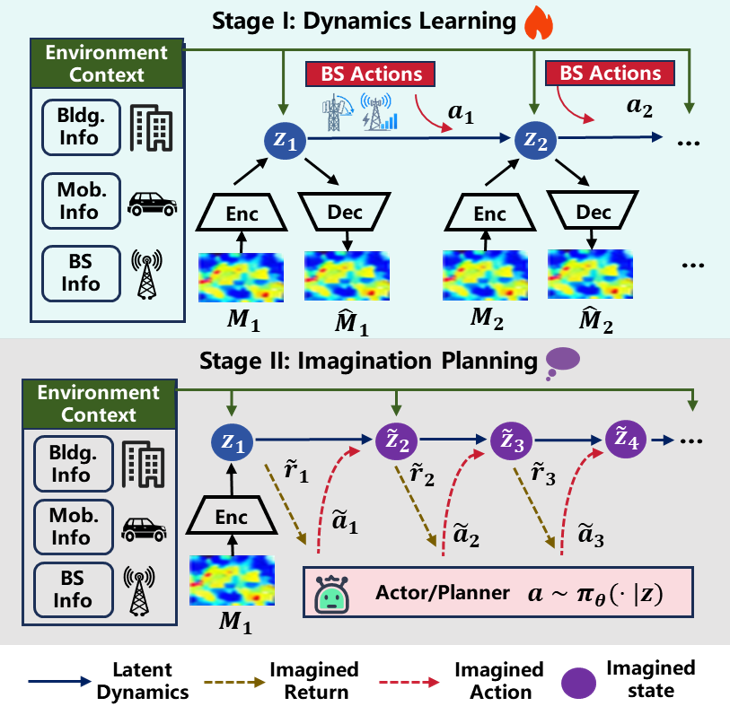}}
\caption{Paradigm of RadioDreamer. The figure illustrates a one-step rollout example, while RadioDreamer performs multi-step rollouts in practice.}
\label{paradigm}
\vspace{-0.2cm}
\end{figure}

To address the challenges above, we propose RadioDreamer, as illustrated in
Fig.~\ref{paradigm}. RadioDreamer follows a two-stage paradigm consisting of
\emph{dynamics learning} and \emph{imagination planning}. The first stage
learns the multi-step evolution of the radio environment conditioned on BS
actions and environmental dynamics, while the second stage exploits the
learned latent dynamics to evaluate and optimize BS actions.

\begin{figure*}[t]
\centerline{\includegraphics[width=.9\linewidth]{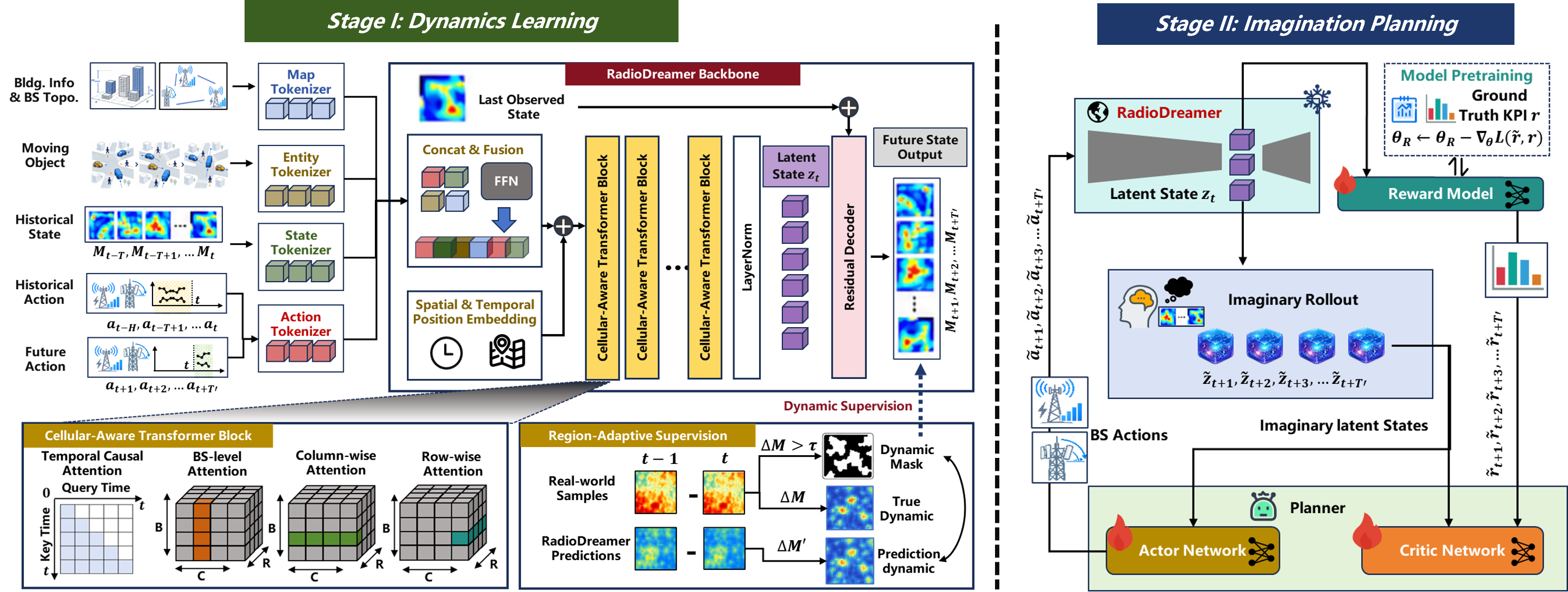}}
\caption{The framework of RadioDreamer. Left: dynamics learning; right:
imagination optimization.}
\label{fig:radiodreamer}
\vspace{-0.2cm}
\end{figure*}

In Stage I, RadioDreamer encodes the historical radio maps $\{\mathbf{M}_1,\mathbf{M}_2,...,\mathbf{M}_t\}$, and reconstructs the radio
map as $\hat{\mathbf{M}}$ through an encoder-decoder architecture:
into latent states:
\begin{equation}
    z_{\kappa}
    =
    \operatorname{Enc}_{\theta}
    \left(
        \mathbf{M}_{\kappa}
    \right),
    \widehat{\mathbf{M}}_{\kappa}
    =
    \operatorname{Dec}_{\theta}
    \left(
        z_{\kappa}
    \right),
    \kappa=1,\ldots,t.
\end{equation}
Given the historical latent states, historical and future BS actions, static
scene, and historical moving-object states, RadioDreamer learns the distribution of the future latent trajectory:
\begin{equation}
\begin{aligned}
    z_{t+1:t+T'}
    \sim
    p_{\theta}
    \Big(
        z_{t+1:t+T'}
        \mid
        z_{t-T:t},
        \mathbf{a}_{t-T:t+T'},
        \mathbf{S},
        \mathbf{N}_{t-T:t}
    \Big).
\end{aligned}
\label{eq:paradigm_transition}
\end{equation}
RadioDreamer therefore learns how a sequence of BS configurations and
environmental dynamics jointly drives the multi-step evolution of the radio
environment, rather than independently mapping each network configuration to
a snapshot.

In Stage II, the learned RadioDreamer is frozen and used as an imagined
environment for network optimization. Starting from the latent state $z_t$
inferred from the latest observation, the Actor/Planner sequentially generates
imagined BS actions $\tilde{\mathbf{a}}_{t+k+1}$, and RadioDreamer rolls out
the corresponding latent states $\tilde{z}_{t+k+1}$. A Reward Model evaluates
each imagined state--action pair and produces the reward
$\tilde{r}_{t+k+1}$. Repeating this process yields a multi-step imagined
state--action--reward trajectory, enabling the policy to optimize the
long-term effects directly in the latent space without
decoding every intermediate state into a full-resolution radio map.

\subsection{Stage I: Dynamics Learning}
\label{subsubsec:dynamics_learning}

The left part of Fig.~\ref{fig:radiodreamer} shows the
dynamics-learning stage. RadioDreamer takes four inputs: the static
scene and BS topology $\mathbf{S}$, historical moving-object states
$\mathbf{N}_{t-T:t}$, historical radio maps $\mathbf{M}_{t-T:t}$, and the BS
action sequence $\mathbf{a}_{t-T:t+T'}$. These inputs are separately tokenized,
fused into a common representation, and fed into the
Transformer-based backbone. 

\paragraph{Multimodal Tokenization and Fusion}

RadioDreamer uses four modality-specific tokenizers to project heterogeneous
radio-environment information into a unified $d$-dimensional representation
space. The Map Tokenizer encodes the static scene $\mathbf{S}$, including
building information and BS topology, as
$\mathbf{X}^{S}=E_{\mathrm{map}}(\mathbf{S})$. The Entity Tokenizer encodes
historical moving-object states as
$\mathbf{X}^{N}=E_{\mathrm{entity}}(\mathbf{N}_{t-T:t})$. The State Tokenizer
partitions each historical radio map into spatial patches and encodes it as
$z_{\tau}=E_{\mathrm{state}}(\mathbf{M}_{\tau})$ for
$\tau=t-T,\ldots,t$. The Action Tokenizer embeds historical and future BS
actions as
$\mathbf{X}^{a}=E_{\mathrm{action}}(\mathbf{a}_{t-T:t+T'})$. Historical
actions provide the control conditions associated with the observed radio-map
sequence, while future actions specify the network configurations under which
future radio states are generated. The modality-specific tokens are
concatenated and projected by a feed-forward fusion module:
\begin{equation}
    \mathbf{X}^{(0)}
    =
    F_{\mathrm{fuse}}
    \left(
        \mathbf{X}^{S},
        \mathbf{X}^{N},
        z_{t-T:t},
        \mathbf{X}^{a}
    \right)
    +
    \mathbf{P}^{\mathrm{tem}}
    +
    \mathbf{P}^{\mathrm{spa}},
    \label{eq:multimodal_fusion}
\end{equation}
where $\mathbf{P}^{\mathrm{tem}}$ and $\mathbf{P}^{\mathrm{spa}}$ denote
learnable absolute temporal and spatial positional embeddings, respectively.
The temporal embedding assigns a distinct representation to each time slot
and is shared across BSs and spatial patches, whereas the spatial embedding
encodes the location of each radio-map patch and is shared across time slots
and BSs, which have been
widely used in Transformer architectures to
distinguish heterogeneous token groups~\cite{devlin2019bert,
dosovitskiy2021image,bertasius2021spacetime}.

\paragraph{Cellular-Aware Transformer Block}



Unlike natural images, radio maps arise from the coupled effects of network
control, cellular topology, and radio propagation. Their evolution is
directional, as future radio states are driven by BS actions and environmental
changes. At each location, received signal and interference depend jointly on
multiple BSs, while propagation continuity, buildings, moving objects, and cell
boundaries produce structured yet spatially non-uniform variations.

To capture these properties, we develop a cellular-aware transformer block
that factorizes token interactions along the temporal, BS, column, and row
dimensions. Let
$\mathbf{X}^{(\ell)}\in
\mathbb{R}^{L\times B\times R'\times C'\times d}$
denote the input to the $\ell$-th block, where $L$ is the sequence length,
$B$ is the number of BSs, $R'\times C'$ is the spatial token grid, and $d$ is
the token dimension.


The temporal causal attention branch models the evolution of the radio environment.
For every BS and spatial location, attention is applied along the temporal
axis as
\begin{equation}
\begin{aligned}
    \mathbf{Y}^{(\ell)}
    =
    \mathbf{X}^{(\ell)}
    +
    \operatorname{Softmax}
    \left(
        \frac{
            \mathbf{Q}_{\mathrm{tem}}
            \mathbf{K}_{\mathrm{tem}}^{\mathsf T}
        }{\sqrt{d_h}}
        +
        \mathbf{\Gamma}
    \right)
    \mathbf{V}_{\mathrm{tem}},
\end{aligned}
\label{eq:temporal_attention}
\end{equation}
where $d_h$ is the dimension of each attention head, and $\mathbf{\Gamma}$ is
the causal mask, with $\Gamma_{u,v}=0$ for $v\leq u$ and
$\Gamma_{u,v}=-\infty$ otherwise.
The causal mask prevents a latent state from accessing radio-state tokens at
later time slots, allowing the model to infer the radio-state trajectory
conditioned on a prescribed action sequence without observing future radio
maps.

The remaining branches characterize the radio map within each time slot.
BS-level attention relates the site-conditioned tokens associated with the
same spatial patch, enabling the model to capture changes in serving-cell
dominance and inter-cell interference caused by power and downtilt
adjustments. Column-wise and row-wise attention model spatial dependencies
along the two axes of the radio map. Their consecutive application allows
information from any spatial patch to reach another patch through the
two-dimensional grid, while avoiding the loss of spatial organization caused
by global token flattening. These operations are written as
\begin{align}
    \mathbf{U}^{(\ell)}_{n,:,i,j,:}
    &=
    \mathbf{Y}^{(\ell)}_{n,:,i,j,:}
    +
    \operatorname{Attn}_{\mathrm{BS}}
    \left(
        \operatorname{LN}
        \left(
            \mathbf{Y}^{(\ell)}_{n,:,i,j,:}
        \right)
    \right),
    \label{eq:bs_attention}\\
    \mathbf{V}^{(\ell)}_{n,b,i,:,:}
    &=
    \mathbf{U}^{(\ell)}_{n,b,i,:,:}
    +
    \operatorname{Attn}_{\mathrm{col}}
    \left(
        \operatorname{LN}
        \left(
            \mathbf{U}^{(\ell)}_{n,b,i,:,:}
        \right)
    \right),
    \label{eq:column_attention}\\
    \mathbf{W}^{(\ell)}_{n,b,:,j,:}
    &=
    \mathbf{V}^{(\ell)}_{n,b,:,j,:}
    +
    \operatorname{Attn}_{\mathrm{row}}
    \left(
        \operatorname{LN}
        \left(
            \mathbf{V}^{(\ell)}_{n,b,:,j,:}
        \right)
    \right).
    \label{eq:row_attention}
\end{align}
Here, $\operatorname{Attn}_{\mathrm{BS}}(\cdot)$,
$\operatorname{Attn}_{\mathrm{col}}(\cdot)$, and
$\operatorname{Attn}_{\mathrm{row}}(\cdot)$ denote multi-head self-attention
applied along the BS, column, and row dimensions, respectively. $\operatorname{LN}(\cdot)$ denotes
Layer Normalization applied along the feature dimension $d$ before attention.

\paragraph{Region-Adaptive Supervision}

Radio-map dynamics are highly non-uniform across space. Most regions may
remain nearly unchanged, while locations near moving objects, coverage
boundaries, and interference-sensitive areas can undergo pronounced
variations. A uniformly averaged loss is easily dominated by static regions
and may cause the model to overlook these sparse but important dynamics.
We therefore introduce \emph{Region-Adaptive Supervision}, which extracts
region-level dynamic cues from temporal variations in each training sequence.
Instead of using predefined dynamic regions, it constructs a sample-dependent
mask that adapts to the evolving radio environment. The mask directs the world
model toward regions with meaningful variations and explicitly supervises
whether their dynamics are accurately reproduced.

As illustrated in Fig.~\ref{fig:radiodreamer}, the dynamic regions are
identified from consecutive ground-truth radio maps. For the $k$-th future
time slot, the ground-truth variation is defined as
\begin{equation}
    \Delta\mathbf{M}_{t+k}
    =
    \left|
        \mathbf{M}_{t+k}
        -
        \mathbf{M}_{t+k-1}
    \right|,
    \qquad k=1,\ldots,T',
    \label{eq:true_dynamic}
\end{equation}
where the absolute value is computed element-wise. A binary dynamic mask is obtained by:
\begin{equation}
    \mathbf{D}_{t+k}
    =
    \mathbb{I}
    \left(
        \Delta\mathbf{M}_{t+k}
        >
        \tau
    \right),
    \label{eq:dynamic_mask}
\end{equation}
where $\tau$ is the dynamic threshold and $\mathbb{I}(\cdot)$ denotes the
indicator function. The mask is generated independently for each sample and
future time slot. It therefore adapts to the changing locations of moving
objects, coverage transitions, and interference fluctuations instead of
assigning a fixed importance to particular spatial regions.
The corresponding variation in the predicted radio maps is calculated as
\begin{equation}
    \Delta\mathbf{M}'_{t+k}
    =
    \left|
        \widehat{\mathbf{M}}_{t+k}
        -
        \widehat{\mathbf{M}}_{t+k-1}
    \right|,
    \label{eq:predicted_dynamic}
\end{equation}
where $\widehat{\mathbf{M}}_{t}=\mathbf{M}_{t}$ for the first prediction step.
The ground-truth and predicted variations describe the spatial extent and
magnitude of radio-map dynamics, respectively. Their discrepancy within the
selected regions is measured by
\begin{equation}
\begin{aligned}
    \mathcal{L}_{\mathrm{dyn}}
    =
    \sum_{k=1}^{T'}
    \omega_k
    \frac{
        \left\|
            \mathbf{D}_{t+k}
            \odot
            \left(
                \Delta\mathbf{M}'_{t+k}
                -
                \Delta\mathbf{M}_{t+k}
            \right)
        \right\|
    }{
        \left\|
            \mathbf{D}_{t+k}
        \right\|+\epsilon
    },
\end{aligned}
\label{eq:dynamic_supervision}
\end{equation}
where $\odot$ denotes element-wise multiplication, $\epsilon$ is a small
constant for numerical stability, and $\omega_k$ is the temporal weight for
the $k$-th prediction step, with
$\sum_{k=1}^{T'}\omega_k=1$. Normalizing by the number of selected elements
prevents samples with larger dynamic regions from dominating the objective.

\paragraph{Dynamics Learning Objective}

The future latent states produced by the RadioDreamer backbone are mapped to the radio-map space through a Residual Decoder. Since large-scale propagation patterns typically evolve slowly, directly estimating absolute radio maps requires repeatedly reconstructing substantial stationary background information. RadioDreamer predicts changes relative to the last observed radio map:
\begin{equation}
    \widehat{\mathbf{M}}_{t+k}
    =
    \mathbf{M}_{t}
    +
    \operatorname{Dec}_{\theta}
    \left(
        z_{t+k}
    \right),
    \qquad
    k=1,\ldots,T',
    \label{eq:residual_decoder}
\end{equation}
where $\operatorname{Dec}_{\theta}(z_{t+k})$ denotes the predicted residual
variation at the $k$-th future time slot. Using $\mathbf{M}_{t}$ as a common
reference preserves the propagation background contained in the latest
observation, while the decoder focuses on the cumulative changes induced over
the forecasting horizon.
The decoded radio maps are supervised by a global prediction objective:
\begin{equation}
    \mathcal{L}_{\mathrm{pred}}
    =
    \sum_{k=1}^{T'}
    \omega_k
    \ell
    \left(
        \widehat{\mathbf{M}}_{t+k},
        \mathbf{M}_{t+k}
    \right),
    \label{eq:prediction_loss}
\end{equation}
where $\ell(\cdot,\cdot)$ measures the discrepancy between the predicted and
ground-truth radio maps. The temporal coefficient $\omega_k$ controls
the contribution of each forecasting step and satisfies
$\sum_{k=1}^{T'}\omega_k=1$.
In summary, the objective for Stage I yields:
\begin{equation}
    \mathcal{L}_{\mathrm{WM}}
    =
    \mathcal{L}_{\mathrm{pred}}
    +
    \lambda_{\mathrm{dyn}}
    \mathcal{L}_{\mathrm{dyn}},
    \label{eq:world_model_objective}
\end{equation}
where $\lambda_{\mathrm{dyn}}$ controls the strength of dynamic-region
supervision. 
The joint objective preserves the propagation features while
capturing how BS reconfiguration and moving objects alter signal strength, thereby improving forecasting capability in
dynamic regions over multi-step rollouts.

\subsection{Imagination Planning}
\label{subsubsec:imagination_planning}


The core of a world model is its ability to learn action-conditioned latent dynamics. Rather than forecasting radio maps under fixed actions, RadioDreamer forms a closed-loop imagined process in which each BS action updates the latent radio state, which then guides the next action, as shown in the right part of Fig.~\ref{fig:radiodreamer}. The Planner can thus roll out alternative action sequences and optimize their cumulative effects over the planning horizon, considering both immediate radio-environment changes and their impact on subsequent states and decisions. Operating directly in the latent space avoids repeated interaction with the physical network and decoding every intermediate state into a full radio map, enabling efficient long-term network optimization before deployment.

\paragraph{Reward Model Pretraining.}

RadioDreamer learns how the radio environment evolves under BS actions, but
its latent states do not directly indicate the resulting network performance.
We therefore train a Reward Model to map the learned radio representation to
a KPI-based reward signal. For a real radio map $\mathbf{M}_{t+k}$ and its BS action $\mathbf{a}_{t+k}$, the ground-truth reward is defined
as $r_{t+k}
    =
    \mathcal{R}
    \left(
        \mathbf{M}_{t+k},
        \mathbf{a}_{t+k}
    \right)$,
where $\mathcal{R}(\cdot)$ specifies the optimization objective, such as coverage, cell-edge performance,
and BS operating cost. Given the latent state
$z_{t+k}=\operatorname{Enc}_{\theta}(\mathbf{M}_{t+k})$, the Reward Model predicts
$ \widehat{r}_{t+k}
    =
    R_{\theta_R}
    \left(
        z_{t+k},
        \mathbf{a}_{t+k}
    \right).$
Its parameters are learned by:
\begin{equation}
    min\mathcal{L}_{R}
    =
    \mathbb{E}_{t,k}
    \left[
        \ell_R
        \left(
            \widehat{r}_{t+k},
            r_{t+k}
        \right)
    \right].
    \label{eq:reward_model_loss}
\end{equation}
The Reward Model thus connects the learned radio dynamics with the network
optimization objective. It allows imagined states to be evaluated directly
in terms of network utility, without decoding every intermediate latent state
into a full-resolution radio map, following the schemes adopted by PlaNet and Dreamer~\cite{hafner2019planet,hafner2020dreamer}. During imagination planning, its parameters are fixed together with RadioDreamer.

\paragraph{Latent Imagination Rollout}

RadioDreamer performs imagination by recursively applying the learned latent
dynamics from the latest observed radio state. Specifically, the current
radio map is encoded as the initial state 
\begin{equation}
    \tilde{z}_{t}
    =
    z_t
    =
    \operatorname{Enc}_{\theta}
    \left(
        \mathbf{M}_{t}
    \right).
    \label{eq:initial_imagined_state}
\end{equation}
Starting from $\tilde{z}_t$, the Actor of planner generates a BS action, and the frozen
RadioDreamer recursively predicts the corresponding latent-state transition:
\begin{align}
    \tilde{\mathbf{a}}_{t+k+1}
    &\sim
    \pi_{\phi}
    \left(
        \cdot
        \mid
        \tilde{z}_{t+k}
    \right),
    \label{eq:imagined_action}\\
    \tilde{z}_{t+k+1}
    &\sim
    p_{\theta}
    \left(
        \cdot
        \mid
        \tilde{z}_{t+k},
        \tilde{\mathbf{a}}_{t+k+1},
        \mathbf{S},
        \mathbf{N}_{t-T:t}
    \right),
    \label{eq:imagined_latent_transition}\\
    \tilde{r}_{t+k+1}
    &=
    R_{\theta_R}
    \left(
        \tilde{z}_{t+k+1},
        \tilde{\mathbf{a}}_{t+k+1}
    \right),
    \
    k=0,\ldots,T'-1.
    \label{eq:imagined_reward}
\end{align}
Each predicted latent state is then used by the Actor to select the next BS
action, yielding an imagined closed-loop \{state, action, reward\} trajectory:
\begin{equation}
    \widetilde{\mathcal{T}}_{t}
    =
    \left\{
        \left(
            \tilde{z}_{t+k},
            \tilde{\mathbf{a}}_{t+k+1},
            \tilde{r}_{t+k+1}
        \right)
    \right\}_{k=0}^{T'-1}.
    \label{eq:imagined_trajectory}
\end{equation}

By sampling alternative actions from the same latent state, RadioDreamer
generates counterfactual trajectories for different power and downtilt
strategies. These trajectories are rolled out directly in the latent space,
avoiding physical-network interaction and intermediate radio-map decoding, and
provide imagined experience for long-term policy optimization.

\paragraph{Policy Learning from Imagination}
During this stage, RadioDreamer and the Reward Model
remain fixed, and only the Actor and Critic are updated.
The imagined trajectory $\widetilde{\mathcal{T}}_{t}$ provides the state
transitions and rewards required for policy learning. The Critic estimates
the expected long-term return from each imagined latent state:
\begin{equation}
    \tilde{V}_{t+k}
    =
    V_{\psi}
    \left(
        \tilde{z}_{t+k}
    \right),
    \label{eq:imagined_value}
\end{equation}
where $\psi$ denotes the Critic parameters. To incorporate rewards accumulated
over the rollout horizon, we construct the learning target using the
$\lambda$-return:
\begin{equation}
    G_{t+k}^{\lambda}
    =
    \tilde{r}_{t+k+1}
    +
    \gamma
    \Big[
        (1-\lambda)
        V_{\bar{\psi}}
        \left(
            \tilde{z}_{t+k+1}
        \right)
        +
        \lambda
        G_{t+k+1}^{\lambda}
    \Big],
\label{eq:lambda_return}
\end{equation}
where $\gamma$ is the discount factor and $V_{\bar{\psi}}$ denotes the target
Critic. 

The Critic learns the long-term value of each imagined latent state by
approximating its corresponding $\lambda$-return. The Actor, in turn, learns
to generate BS actions that maximize the cumulative reward along the imagined
trajectory, with an entropy term retained to encourage exploration among
alternative power and downtilt configurations. Their optimization objectives
are given by
\begin{subequations}
\label{eq:actor_critic_objectives}
\begin{align}
    \mathcal{L}_{\mathrm{critic}}
    &=
    \frac{1}{T'}
    \sum_{k=0}^{T'-1}
    \ell_V
    \left(
        V_{\psi}
        \left(
            \tilde{z}_{t+k}
        \right),
        G_{t+k}^{\lambda}
    \right),
    \label{eq:critic_loss}\\
    \mathcal{L}_{\mathrm{actor}}
    &=
    -
    \frac{1}{T'}
    \sum_{k=0}^{T'-1}
    \mathbb{E}
    \left[
        G_{t+k}^{\lambda}
        +
        \beta
        \mathcal{H}
        \left(
            \pi_{\phi}
            \left(
                \cdot
                \mid
                \tilde{z}_{t+k}
            \right)
        \right)
    \right],
    \label{eq:actor_loss}
\end{align}
\end{subequations}
where $\mathcal{H}(\cdot)$ denotes the policy entropy and $\beta$ controls
the exploration strength.


\section{Evaluation}
\label{sec:evaluation}

We evaluate RadioDreamer on a simulated dataset of over 400,000 radio-map frames to answer three research questions:

\begin{itemize}
    \item \textbf{RQ1:} Can RadioDreamer reliably predict radio-map evolution
    in dynamic radio environments?
    \item \textbf{RQ2:} Can RadioDreamer model action-state dynamics and
    generalize to unseen BS configurations?
    \item \textbf{RQ3:} How effective is RadioDreamer's imagination-based
    planning for BS-control optimization?
\end{itemize}




\subsection{Evaluation Setup}
\label{subsec:evaluation_setup}

\paragraph{Scenario and Dataset Construction}



To the best of our known, no publicly available real-world dataset
provides spatio-temporal radio maps synchronized with time-varying BS parameters, particularly transmit power and electrical downtilt.
We therefore construct a controlled simulation dataset using Sionna RT, in
which BS actions, environmental dynamics, and the resulting RSRP and SINR
maps are temporally aligned.
The dataset covers 3 urban scenes: Street Canyon, Etoile, and Munich.
Each scene corresponds to a fixed urban map with a distinct building layout
and propagation contexts. The data generation, model training, and evaluation are
performed independently for each scene without cross-scene joint training.
In each scene, the evaluated region covers
$120\,\mathrm{m}\times80\,\mathrm{m}$ and is discretized at a spatial
resolution of $2.5$~m, resulting in a $48\times32$ radio map. The carrier frequency is set to
$3.5$~GHz, and Ray tracing
considers line-of-sight and specular-reflection paths with a maximum
propagation depth of 4.

We define an episode as a sequence of 32 consecutive radio-environment
frames generated under a sampled BS deployment and a set of moving-object
trajectories. Each episode contains 100 moving objects and 2--6 active BSs,
with transmit powers ranging from 10 to 30~dBm and electrical downtilts
selected from
$\{0^\circ,3.75^\circ,7.5^\circ,11.25^\circ,15^\circ\}$.

For each episode, a complete 32-frame trajectory of moving objects is independently
generated using a unique random seed. The parameterized analytical generator
combines directional movement, periodic boundary re-entry, stop-and-go motion,
lateral lane-change-like variations, and irregular two-dimensional motion.
Initial positions, directions, speeds, amplitudes, and phases of moving objects are
independently sampled according to the motion distribution specified by each
evaluation protocol.
Random seeds are partitioned into mutually
exclusive training, validation, and test sets before data generation. Thus, no
trajectory instance or overlapping trajectory segment is shared across
different data splits. For each scene, we generate 3,600 training episodes,
480 validation episodes, and 960 test episodes. Across the three scenes, the
resulting dataset contains 15,120 episodes and 483,840 temporally ordered
radio-environment frames.
Given the first 16 frames of radio maps, moving object states, BS parameters, and
the urban layout, the models predict the subsequent 16 RSRP and SINR maps
without observing future object states. Forecasting performance is evaluated
at horizons $H\in\{1,8,16\}$. All experiments are conducted on an Intel Core
i7-14700KF CPU, 32~GB of memory, and an NVIDIA GeForce RTX 5070 Ti GPU with
16~GB of memory.
We consider the following three evaluation
protocols:

\begin{table*}[ht]
    \centering
    \fontsize{7.5pt}{8.5pt}\selectfont

    \newcommand{\bestresult}[1]{%
        \cellcolor{orange!25}$\mathbf{#1}$%
    }
    \newcommand{\secondresult}[1]{%
        \cellcolor{cyan!20}$\underline{#1}$%
    }

    \caption{
    Radio map forecasting performance of full regions.
    Results are reported as mean $\pm$ standard deviation
    in dB, where lower values indicate better performance.
    (\colorbox{orange!25}{\textbf{Bold}} indicates the best result and
    \colorbox{cyan!20}{\underline{underlined}} indicates the second-best result.)
    }
    \label{tab:fmae_forecasting_performance}

    \setlength{\tabcolsep}{3.8pt}
    \renewcommand{\arraystretch}{1.10}

    \resizebox{\textwidth}{!}{%
    \begin{tabular}{@{}lc*{9}{c}@{}}
        \toprule

\multirow{2}{*}{\textbf{Method}}
& \multirow[c]{2}{*}{%
    \shortstack[c]{\textbf{Radio}\\[-1pt]\textbf{Quantity}}%
}
& \multicolumn{3}{c}{\textbf{Street Canyon}}
& \multicolumn{3}{c}{\textbf{Etoile}}
& \multicolumn{3}{c}{\textbf{Munich}} \\

        \cmidrule(lr){3-5}
        \cmidrule(lr){6-8}
        \cmidrule(lr){9-11}

        & & F-MAE (H=1) & F-MAE (H=8) & F-MAE (H=16)
        &  F-MAE (H=1) & F-MAE (H=8) & F-MAE (H=16)
        &  F-MAE (H=1) & F-MAE (H=8) & F-MAE (H=16)\\

        \midrule

        \multirow{2}{*}{3GPP UMi}
        & RSRP
        & $11.115 \pm 0.153$
        & $11.069 \pm 0.161$
        & $11.101 \pm 0.117$
        & $2.963 \pm 0.021$
        & $2.953 \pm 0.016$
        & $2.955 \pm 0.014$
        & $7.824 \pm 0.227$
        & $7.817 \pm 0.228$
        & $7.783 \pm 0.258$ \\

        & SINR
        & $28.661 \pm 0.256$
        & $28.757 \pm 0.223$
        & $28.794 \pm 0.273$
        & $2.134 \pm 0.089$
        & $2.140 \pm 0.091$
        & $2.145 \pm 0.079$
        & $23.908 \pm 0.398$
        & $23.814 \pm 0.500$
        & $23.867 \pm 0.449$ \\

        \midrule

        \multirow{2}{*}{RadioUNet}
        & RSRP
        & $1.334 \pm 0.576$
        & $1.373 \pm 0.568$
        & $1.372 \pm 0.565$
        & $0.829 \pm 0.200$
        & $0.886 \pm 0.195$
        & $0.888 \pm 0.193$
        & $3.712 \pm 3.761$
        & $3.700 \pm 3.683$
        & $3.735 \pm 3.723$ \\

        & SINR
        & $1.590 \pm 0.755$
        & $1.625 \pm 0.754$
        & $1.619 \pm 0.749$
        & $1.079 \pm 0.349$
        & \secondresult{1.101 \pm 0.329}
        & \secondresult{1.111 \pm 0.334}
        & $4.809 \pm 4.851$
        & $4.806 \pm 4.803$
        & $4.862 \pm 4.859$ \\

        \midrule

        \multirow{2}{*}{PMNet}
        & RSRP
        & $1.177 \pm 0.540$
        & \secondresult{1.179 \pm 0.538}
        & \secondresult{1.173 \pm 0.532}
        & $0.829 \pm 0.212$
        & \secondresult{0.877 \pm 0.208}
        & \secondresult{0.879 \pm 0.206}
        & $3.646 \pm 3.781$
        & $3.603 \pm 3.721$
        & $3.614 \pm 3.734$ \\

        & SINR
        & $1.443 \pm 0.674$
        & \secondresult{1.450 \pm 0.674}
        & \secondresult{1.441 \pm 0.669}
        & $1.099 \pm 0.365$
        & $1.109 \pm 0.348$
        & $1.119 \pm 0.354$
        & $4.799 \pm 4.897$
        & $4.754 \pm 4.839$
        & $4.803 \pm 4.891$ \\

        \midrule

        \multirow{2}{*}{Pathloss-PINN}
        & RSRP
        & $4.798 \pm 0.824$
        & $4.823 \pm 0.806$
        & $4.819 \pm 0.801$
        & $1.244 \pm 0.012$
        & $1.296 \pm 0.009$
        & $1.295 \pm 0.005$
        & $4.588 \pm 1.593$
        & $4.578 \pm 1.542$
        & $4.573 \pm 1.569$ \\

        & SINR
        & $10.597 \pm 0.857$
        & $10.632 \pm 0.884$
        & $10.615 \pm 0.872$
        & $1.745 \pm 0.052$
        & $1.749 \pm 0.051$
        & $1.752 \pm 0.048$
        & $7.932 \pm 2.095$
        & $7.911 \pm 2.072$
        & $7.932 \pm 2.097$ \\

        \midrule

        \multirow{2}{*}{IPP-Net}
        & RSRP
        & $1.607 \pm 0.502$
        & $1.645 \pm 0.498$
        & $1.642 \pm 0.494$
        & \secondresult{0.825 \pm 0.186}
        & $0.888 \pm 0.177$
        & $0.891 \pm 0.176$
        & $3.812 \pm 3.674$
        & $3.796 \pm 3.603$
        & $3.815 \pm 3.625$ \\

        & SINR
        & $1.893 \pm 0.661$
        & $1.927 \pm 0.664$
        & $1.921 \pm 0.656$
        & \secondresult{1.063 \pm 0.335}
        & $1.110 \pm 0.304$
        & $1.120 \pm 0.309$
        & $4.973 \pm 4.805$
        & $4.967 \pm 4.768$
        & $5.007 \pm 4.808$ \\

        \midrule

        \multirow{2}{*}{Height-RM-CNN}
        & RSRP
        & $1.223 \pm 0.606$
        & $1.270 \pm 0.597$
        & $1.267 \pm 0.592$
        & $0.839 \pm 0.216$
        & $0.890 \pm 0.212$
        & $0.892 \pm 0.210$
        & $4.330 \pm 4.482$
        & $4.312 \pm 4.413$
        & $4.328 \pm 4.430$ \\

        & SINR
        & $1.454 \pm 0.742$
        & $1.494 \pm 0.737$
        & $1.491 \pm 0.736$
        & $1.125 \pm 0.366$
        & $1.131 \pm 0.351$
        & $1.140 \pm 0.356$
        & $5.683 \pm 5.861$
        & $5.675 \pm 5.825$
        & $5.725 \pm 5.874$ \\

        \midrule

        \multirow{2}{*}{Dreamer-RSSM}
        & RSRP
        & \secondresult{1.142 \pm 0.030}
        & $1.314 \pm 0.038$
        & $1.322 \pm 0.037$
        & $0.850 \pm 0.006$
        & $0.958 \pm 0.012$
        & $0.971 \pm 0.009$
        & \secondresult{1.520 \pm 0.301}
        & \secondresult{1.692 \pm 0.342}
        & \secondresult{1.736 \pm 0.364} \\

        & SINR
        & \secondresult{1.414 \pm 0.015}
        & $1.675 \pm 0.024$
        & $1.694 \pm 0.024$
        & $1.064 \pm 0.022$
        & $1.245 \pm 0.031$
        & $1.261 \pm 0.025$
        & \secondresult{1.950 \pm 0.306}
        & \secondresult{2.192 \pm 0.345}
        & \secondresult{2.244 \pm 0.367} \\

        \midrule

        \multirow{2}{*}{\textbf{RadioDreamer (Ours)}}
        & \textbf{RSRP}
        & \bestresult{0.905 \pm 0.151}
        & \bestresult{0.894 \pm 0.157}
        & \bestresult{0.893 \pm 0.154}
        & \bestresult{0.620 \pm 0.013}
        & \bestresult{0.623 \pm 0.017}
        & \bestresult{0.623 \pm 0.015}
        & \bestresult{1.244 \pm 0.418}
        & \bestresult{1.176 \pm 0.423}
        & \bestresult{1.178 \pm 0.429} \\

        & \textbf{SINR}
        & \bestresult{1.081 \pm 0.166}
        & \bestresult{1.068 \pm 0.175}
        & \bestresult{1.066 \pm 0.174}
        & \bestresult{0.894 \pm 0.027}
        & \bestresult{0.896 \pm 0.031}
        & \bestresult{0.894 \pm 0.028}
        & \bestresult{1.525 \pm 0.429}
        & \bestresult{1.446 \pm 0.442}
        & \bestresult{1.449 \pm 0.443} \\

        \bottomrule
    \end{tabular}%
    }
\end{table*}

$\bullet$ \textbf{T1:}
The training and test episodes follow the same motion-generation rules,
parameter ranges, and candidate BS-position set, but use mutually exclusive
trajectory seeds. Their BS deployments and action sequences are also sampled
independently. This protocol evaluates forecasting on unseen trajectory
instances drawn from the training distribution.


$\bullet$ \textbf{T2:}
The test episodes use BS coordinates excluded from training. The
trajectories of moving objects follow the same motion-generation rules and parameter ranges as
the training data but are generated using independent trajectory seeds. This
protocol evaluates whether RadioDreamer can transfer the learned propagation
dynamics to unseen BS deployments.

$\bullet$ \textbf{T3:}
The test episodes use both BS coordinates and mobility regimes excluded from
training. The held-out mobility regimes contain motion-generation settings or
parameter ranges not observed during training. This protocol evaluates
RadioDreamer when both the network deployment and moving object dynamics
differ from those in the training data.

\paragraph{Baselines}
The baselines are organized into three categories:
(i) \textbf{Statistical model.}
We adopt the 3GPP TR~38.901 Urban Micro Street Canyon
(3GPP-UMi) model as a non-learning baseline
\cite{3gpp38901}. It estimates radio propagation using standardized
distance-dependent LOS and NLOS path-loss formulations.
(ii) \textbf{Static radio-map models.}
RadioUNet uses a U-Net-based encoder--decoder to infer radio maps
from environmental geometry and transmitter information
\cite{levie2021radiounet}. PMNet introduces residual blocks and dilated
convolutions to capture multi-scale propagation patterns
\cite{lee2023pmnet}. Pathloss-PINN incorporates physical dependencies of the
spatial path-loss field into the learning objective
\cite{limmer2023pathloss}. IPP-Net combines a U-Net architecture with
ray-tracing data and a 3GPP propagation model to improve generalization
\cite{feng2025ippnet}. Height-RM-CNN adopts the height-aware CNN to guide the radio map prediction
\cite{jaensch2026radiomap}.
(iii) \textbf{World model.}
We include Dreamer-RSSM as a general-purpose world-model baseline
\cite{hafner2020dreamer}. It uses a recurrent state-space
model to learn stochastic latent dynamics, allowing us to distinguish the
benefit of RadioDreamer's radio-specific architecture from that of a generic
world model.

\paragraph{Metrics}
We evaluate forecasting performance using 3 metrics:
\textbf{(i) Full-map Mean Absolute Error (F-MAE)} measures the average error over the
entire radio map and is calculated as
$\mathrm{F\text{-}MAE}_{H}
=
\mathbb{E}_{x}
\left|
\widehat{\mathbf{M}}_{t+H}(x)
-
\mathbf{M}_{t+H}(x)
\right|$.
\textbf{(ii) Dynamic-region Mean Absolute Error (D-MAE)} measures the prediction error
over the regions selected by the dynamic mask $\mathbf{D}_{t+H}$ defined in
Eq.~\eqref{eq:dynamic_mask}:
$\mathrm{D\text{-}MAE}_{H}
=
\mathbb{E}_{x}\left|
\mathbf{D}_{t+H}(x)[
\widehat{\mathbf{M}}_{t+H}(x)
-
\mathbf{M}_{t+H}(x)]
\right|$.
\textbf{(iii) Signal Direction (SD)} is adapted from the directional-accuracy metric commonly
used in time-series forecasting~\cite{pesaran1992simple}. It measures whether
the predicted radio-map variation has the same direction as the ground-truth
variation. For a spatial location $x$ and rollout horizon $H$, we define
$\Delta\mathbf{M}_{H}(x)
=\mathbf{M}_{t+H}(x)-\mathbf{M}_{t}(x)$ and
$\Delta\widehat{\mathbf{M}}_{H}(x)
=\widehat{\mathbf{M}}_{t+H}(x)-\mathbf{M}_{t}(x)$.
Accordingly,
\begin{equation}
    \mathrm{SD}_{H}
    =
    \mathbb{E}_{x}
    \mathbb{I}
    \left(
        \operatorname{sign}
        \big(\Delta\widehat{\mathbf{M}}_{H}(x)\big)
        =
        \operatorname{sign}
        \big(\Delta\mathbf{M}_{H}(x)\big)
    \right).
    \label{eq:sd}
\end{equation}
Here, $\operatorname{sign}(\cdot)$ indicates whether the radio state increases,
decreases, or remains unchanged. A higher SD indicates more reliable
prediction of the radio-state evolution direction.

\begin{table*}[t]
    \centering
    \fontsize{6pt}{7pt}\selectfont

    \newcommand{\bestresult}[1]{%
        \cellcolor{orange!25}$\mathbf{#1}$%
    }
    \newcommand{\secondresult}[1]{%
        \cellcolor{cyan!20}$\underline{#1}$%
    }

    \caption{
    Radio map forecasting performance under dynamic regions.  Results are reported as mean $\pm$ standard deviation
    in dB.
    }
    \label{tab:a_d1_comparison}

    \setlength{\tabcolsep}{3.8pt}
    \renewcommand{\arraystretch}{1}

    \begin{tabular*}{\textwidth}{
        @{\extracolsep{\fill}}lc*{6}{c}@{}
    }
        \toprule

\multirow{2}{*}{\textbf{Method}}
& \multirow[c]{2}{*}{%
    \shortstack[c]{\textbf{Radio}\\[-1pt]\textbf{Quantity}}%
}
& \multicolumn{2}{c}{\textbf{Street Canyon}}
& \multicolumn{2}{c}{\textbf{Etoile}}
& \multicolumn{2}{c}{\textbf{Munich}} \\

        \cmidrule(lr){3-4}
        \cmidrule(lr){5-6}
        \cmidrule(lr){7-8}

        & & D-MAE (H=8)
        & SD $\uparrow$
        & D-MAE (H=8)
        & SD $\uparrow$
        & D-MAE (H=8)
        & SD $\uparrow$ \\

        \midrule

        \multirow{2}{*}{RadioUNet}
        & RSRP
        & $3.435 \pm 0.379$
        & $45.91 \pm 0.17$
        & $2.359 \pm 0.079$
        & $46.40 \pm 0.13$
        & $7.617 \pm 4.407$
        & $46.07 \pm 0.47$ \\

        & SINR
        & $3.707 \pm 0.547$
        & $47.06 \pm 0.08$
        & $2.351 \pm 0.203$
        & $46.46 \pm 0.34$
        & $8.660 \pm 5.384$
        & $47.56 \pm 0.42$ \\

        \midrule

        \multirow{2}{*}{PMNet}
        & RSRP
        & \secondresult{3.049 \pm 0.361}
        & $49.66 \pm 0.23$
        & $2.280 \pm 0.092$
        & $45.85 \pm 0.19$
        & $7.271 \pm 4.531$
        & \secondresult{49.10 \pm 0.16} \\

        & SINR
        & \secondresult{3.421 \pm 0.483}
        & $49.72 \pm 0.24$
        & $2.319 \pm 0.219$
        & \secondresult{48.91 \pm 0.12}
        & $8.420 \pm 5.452$
        & \secondresult{49.36 \pm 0.31} \\

        \midrule

        \multirow{2}{*}{Pathloss-PINN}
        & RSRP
        & $6.441 \pm 0.863$
        & $25.96 \pm 0.18$
        & $2.437 \pm 0.027$
        & $29.51 \pm 0.12$
        & $7.802 \pm 2.163$
        & $25.43 \pm 0.20$ \\

        & SINR
        & $12.184 \pm 1.103$
        & $23.55 \pm 0.10$
        & $2.732 \pm 0.081$
        & $19.52 \pm 0.14$
        & $10.836 \pm 2.614$
        & $21.13 \pm 0.18$ \\

        \midrule

        \multirow{2}{*}{IPP-Net}
        & RSRP
        & $3.599 \pm 0.374$
        & $46.02 \pm 0.06$
        & $2.405 \pm 0.063$
        & $46.27 \pm 0.14$
        & $7.694 \pm 4.475$
        & $46.51 \pm 0.56$ \\

        & SINR
        & $3.890 \pm 0.528$
        & $47.38 \pm 0.14$
        & $2.420 \pm 0.167$
        & $46.69 \pm 0.42$
        & $8.802 \pm 5.518$
        & $47.99 \pm 0.66$ \\

        \midrule

        \multirow{2}{*}{Height-RM-CNN}
        & RSRP
        & $3.386 \pm 0.366$
        & $45.78 \pm 0.25$
        & \secondresult{2.229 \pm 0.090}
        & \secondresult{48.07 \pm 0.12}
        & $8.339 \pm 5.553$
        & $46.28 \pm 0.48$ \\

        & SINR
        & $3.643 \pm 0.508$
        & $47.05 \pm 0.24$
        & \secondresult{2.313 \pm 0.225}
        & $48.68 \pm 0.22$
        & $9.695 \pm 6.908$
        & $47.65 \pm 0.36$ \\

        \midrule

        \multirow{2}{*}{Dreamer (Soft)}
        & RSRP
        & $4.632 \pm 0.093$
        & \secondresult{51.68 \pm 0.24}
        & $3.156 \pm 0.024$
        & $10.66 \pm 0.19$
        & \secondresult{6.315 \pm 0.468}
        & $7.10 \pm 0.14$ \\

        & SINR
        & $5.008 \pm 0.046$
        & \secondresult{53.60 \pm 0.17}
        & $3.612 \pm 0.086$
        & $26.24 \pm 0.45$
        & \secondresult{6.803 \pm 0.432}
        & $12.50 \pm 0.42$ \\

        \midrule

        \multirow{2}{*}{\textbf{RadioDreamer (Ours)}}
        & \textbf{RSRP}
        & \bestresult{2.705 \pm 0.153}
        & \bestresult{80.38 \pm 0.92}
        & \bestresult{1.798 \pm 0.022}
        & \bestresult{81.20 \pm 0.21}
        & \bestresult{3.460 \pm 0.484}
        & \bestresult{80.36 \pm 1.91} \\

        & \textbf{SINR}
        & \bestresult{2.982 \pm 0.150}
        & \bestresult{79.79 \pm 1.17}
        & \bestresult{2.163 \pm 0.053}
        & \bestresult{80.27 \pm 0.35}
        & \bestresult{3.891 \pm 0.476}
        & \bestresult{79.96 \pm 2.20} \\

        \bottomrule
    \end{tabular*}
\end{table*}

\subsection{Forecasting performance (RQ1)}

We first evaluate radio-map forecasting under the T1 protocol. Given the first
16 radio-environment frames and future BS actions, each model predicts the
subsequent 16 radio maps. Results are reported at
$H\in\{1,8,16\}$ for Street Canyon, Etoile, and Munich.
As shown in Table~\ref{tab:fmae_forecasting_performance}, RadioDreamer achieves
the lowest F-MAE in all 18 combinations of scene, radio quantity, and prediction
horizon. It reduces F-MAE by 15.9\%--35.4\% over the strongest baseline in each
setting, with an average reduction of 25.2\%. The gain is particularly large in
Munich, where the propagation conditions are more complex. At $H=16$,
RadioDreamer reduces the RSRP and SINR errors by 32.1\% and 35.4\%,
respectively, compared with Dreamer-RSSM. Its errors also remain stable across
different horizons, indicating reliable long-term forecasting. Static
radio-map models remain competitive in simpler settings but degrade in complex
scenes because they do not explicitly model the temporal coupling among radio
states, BS actions, and environmental changes. Dreamer-RSSM benefits from
latent dynamics modeling, yet remains inferior, suggesting that generic
recurrent dynamics are insufficient to capture cellular interactions and
structured spatial propagation.

Table~\ref{tab:a_d1_comparison} evaluates dynamically changing regions
at $H=8$. RadioDreamer achieves the best dynamic-region MAE and SD for both
RSRP and SINR in all three scenes, reducing MAE by 6.5\%--45.2\% over the
strongest competitor. It also maintains an SD of approximately 80\%, whereas
the best baselines achieve only 48.07\%--53.60\%. This gap indicates that a low
full-map error does not necessarily imply accurate dynamics: a model may
recover the dominant propagation background while predicting the wrong
direction of local radio changes.

\begin{figure}[t]
\centerline{\includegraphics[width=\linewidth]{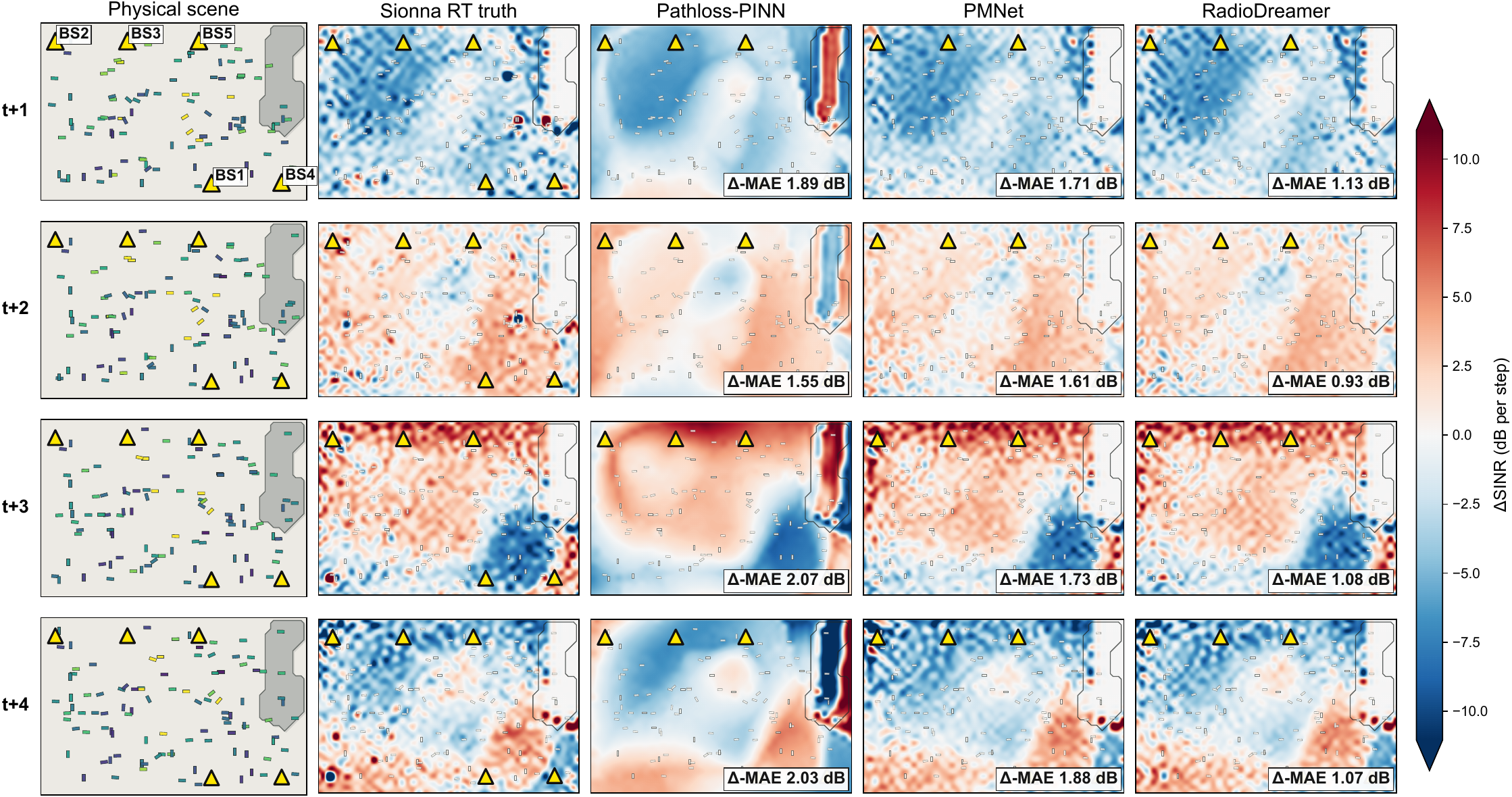}}
\caption{Visualization of Dynamic Radio Map Forecasting Results.}
\label{etoile_dynamic_sinr_t1_t4_scene_baselines_compact}
\end{figure}

The qualitative results in
Fig.~\ref{etoile_dynamic_sinr_t1_t4_scene_baselines_compact} support this
observation. Pathloss-PINN produces mainly smooth large-scale variations and
misses localized changes, while PMNet preserves more spatial detail but still
predicts inaccurate variation patterns. RadioDreamer more closely matches the
Sionna RT ground truth for both positive and negative SINR changes, achieving
$\Delta$-MAE values of 0.93--1.13~dB, compared with 1.55--2.07~dB for
Pathloss-PINN and 1.61--1.88~dB for PMNet. These results show that
RadioDreamer captures both the overall radio-map distribution and the localized
evolution induced by BS actions and moving objects.

\subsection{Action-state Dynamics and Generalization (RQ2)}
\label{subsec:rollout_generalization}

\begin{table}[ht]
    \centering
    \fontsize{7.5pt}{8.5pt}\selectfont

    \newcommand{\bestresult}[1]{%
        \cellcolor{orange!25}$\mathbf{#1}$%
    }
    \newcommand{\secondresult}[1]{%
        \cellcolor{cyan!20}$\underline{#1}$%
    }

    \caption{
Generalization performance measured by F-MAE (H=8)
    }
    \label{tab:sinr_ood_performance}

    \setlength{\tabcolsep}{5.0pt}
    \renewcommand{\arraystretch}{1.10}

    \resizebox{\linewidth}{!}{%
    \begin{tabular}{@{}l*{6}{c}@{}}
        \toprule

        \multirow{2}{*}{\textbf{Method}}
        & \multicolumn{2}{c}{\textbf{Street Canyon}}
        & \multicolumn{2}{c}{\textbf{Etoile}}
        & \multicolumn{2}{c}{\textbf{Munich}} \\

        \cmidrule(lr){2-3}
        \cmidrule(lr){4-5}
        \cmidrule(lr){6-7}

        & \textbf{T2}
        & \textbf{T3}
        & \textbf{T2}
        & \textbf{T3}
        & \textbf{T2}
        & \textbf{T3} \\

        \midrule

        RadioUNet
        & $2.634 \pm 1.444$
        & $2.458 \pm 1.298$
        & $1.456 \pm 0.932$
        & $1.548 \pm 1.051$
        & $10.327 \pm 10.924$
        & $11.037 \pm 11.232$ \\

        \midrule

        PMNet
        & $2.339 \pm 1.160$
        & $2.208 \pm 1.105$
        & $1.487 \pm 0.946$
        & $1.579 \pm 1.062$
        & $10.354 \pm 10.634$
        & $10.997 \pm 10.789$ \\

        \midrule

        Pathloss-PINN
        & $11.623 \pm 4.330$
        & $11.703 \pm 4.360$
        & $1.750 \pm 1.140$
        & $1.795 \pm 1.254$
        & $10.179 \pm 7.900$
        & $10.687 \pm 8.069$ \\

        \midrule

        IPP-Net
        & $2.814 \pm 1.241$
        & $2.660 \pm 1.102$
        & $1.444 \pm 0.965$
        & $1.518 \pm 1.054$
        & $10.411 \pm 10.568$
        & $11.185 \pm 10.778$ \\

        \midrule

        Height-RM-CNN
        & $2.462 \pm 1.222$
        & $2.330 \pm 1.156$
        & $1.518 \pm 0.992$
        & $1.602 \pm 1.094$
        & $12.386 \pm 10.919$
        & $13.217 \pm 10.772$ \\

        \midrule

        Dreamer-RSSM
        & \secondresult{1.701 \pm 0.428}
        & \secondresult{1.684 \pm 0.465}
        & \secondresult{1.233 \pm 0.656}
        & \secondresult{1.251 \pm 0.704}
        & \secondresult{2.605 \pm 1.215}
        & \secondresult{2.624 \pm 1.078} \\

        \midrule

        \textbf{RadioDreamer}
        & \bestresult{1.538 \pm 0.595}
        & \bestresult{1.474 \pm 0.572}
        & \bestresult{0.901 \pm 0.539}
        & \bestresult{0.945 \pm 0.616}
        & \bestresult{1.943 \pm 0.953}
        & \bestresult{2.029 \pm 0.965} \\

        \bottomrule
    \end{tabular}%
    }
\end{table}

\begin{figure}[ht]
\centerline{\includegraphics[width=\linewidth]{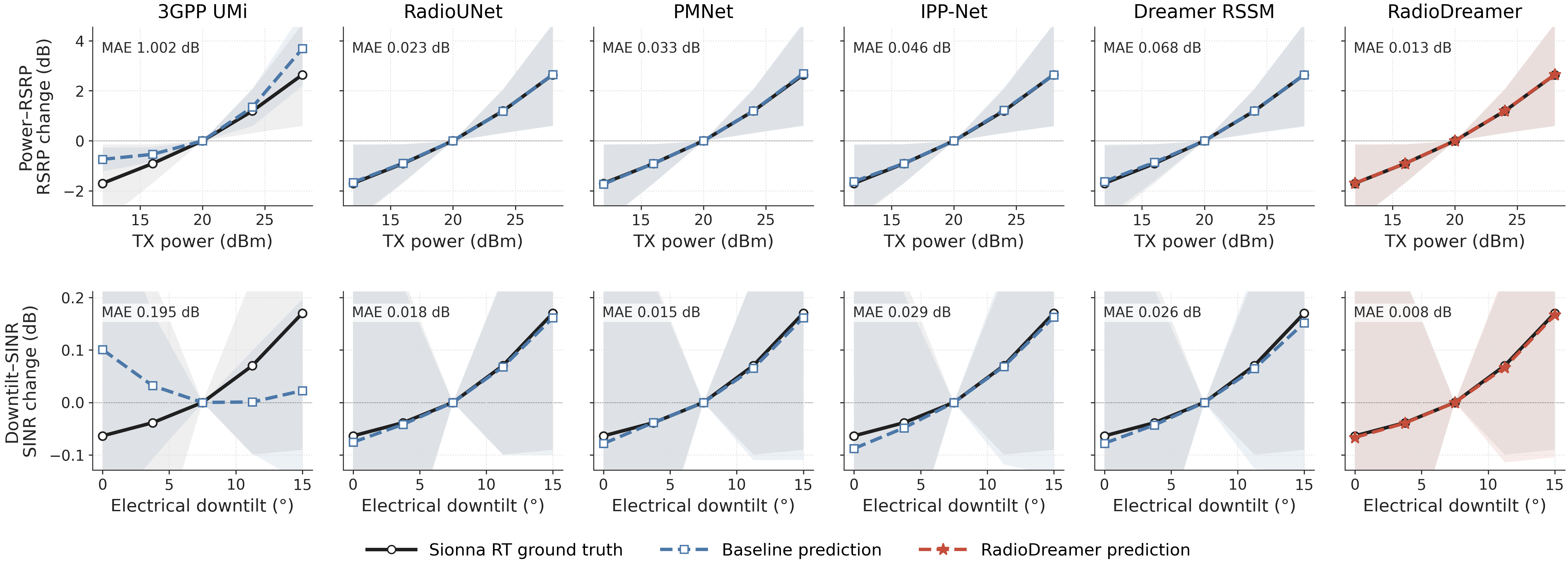}}
\caption{Action--State Intervention Response Results.}
\label{fig:action_state_response}
\vspace{-0.2cm}
\end{figure}

Fig.~\ref{fig:action_state_response} reports the paired intervention results at
$H=8$. RadioDreamer most closely matches the Sionna RT dose--response curves
for both Power--RSRP and Downtilt--SINR. It achieves MAEs of $0.013$~dB and
$0.008$~dB, reducing the errors by 43.5\% and 46.7\% over the strongest
baselines, respectively. In contrast, 3GPP UMi substantially misestimates the
response magnitude and even predicts an incorrect SINR-change direction at
small downtilts. These results show that RadioDreamer accurately captures the
direction, magnitude, and nonlinear response of radio-state changes induced by
BS actions, rather than merely reconstructing static radio maps.

Table~\ref{tab:sinr_ood_performance} reports the generalization results at
$H=8$. RadioDreamer achieves the lowest F-MAE in all six T2 and T3 settings,
reducing the error by 9.6\%--26.9\% over Dreamer-RSSM, with an average reduction
of 20.3\%. The gain is largest in Munich, where static radio-map models produce
errors of approximately 10--13~dB, whereas RadioDreamer achieves 1.943~dB under
T2 and 2.029~dB under T3. Its comparable performance under T2 and T3 further
indicates robustness to jointly held-out BS positions and unseen trajectory
instances.
These results show the structured dynamics
modeling capability of RadioDreamer, which captures temporal action--state dependencies, inter-BS
interactions, and spatial propagation, which enables the learned
dynamics to generalize beyond training configurations and limits error
accumulation during rollout.

\subsection{Imagination Optimization (RQ3)}
\label{subsec:imagination_optimization}

We consider a dynamic BS-configuration problem in the Street Canyon scene,
which contains 100 moving objects and 2--6 active BSs. As vehicles and
pedestrians move through the scene, their time-varying blockage and reflection
effects continuously reshape the coverage and interference distributions. At
each time slot, the controller adjusts the transmit power and electrical
downtilt of every active BS. The objective is to determine a 16-step action
sequence that improves overall coverage and cell-edge signal quality over the
evolving radio environment, rather than optimizing an individual radio-map
snapshot.
The instantaneous reward yields:
\begin{equation}
    r_t
    =
     C_t^{\mathrm{RSRP}}
    +
    0.8 C_t^{\mathrm{SINR}}
    +
    0.35 C_t^{\mathrm{Edge}},
    \label{eq:planning_reward}
\end{equation}
where $C_t^{\mathrm{RSRP}}$ and $C_t^{\mathrm{SINR}}$ denote the proportions
of valid locations with RSRP no lower than $-100$~dBm and SINR no lower than
$0$~dB, respectively.
Let $\Omega_{5\%}^{\mathrm{ref}}$ denote the grid cells with the lowest 5\%
RSRP values in the radio map.
The cell-edge coverage at time slot $t$, denoted by
$C_t^{\mathrm{Edge}}$, is computed as the proportion of locations in the
reference 5\%-worst region $\Omega_{5\%}^{\mathrm{ref}}$ whose RSRP exceeds
$-55$~dBm.
This
objective encourages the policy to preserve broad signal coverage while
improving interference-limited and weak-coverage regions throughout the
16-step planning horizon.
The  optimization problem yields:
\begin{equation}
\begin{aligned}
    \max_{\pi_{\phi}}\quad
    &\mathbb{E}_{\pi_{\phi}}
    \left[
        \sum_{k=0}^{15}
        0.98^{k} r_{t+k}
    \right] \\
    \mathrm{s.t.}\quad
    &10 \leq p_{t+k}^{b}\leq30~\mathrm{dBm},\\
    &\delta_{t+k}^{b}
    \in
    \{0^\circ,3.75^\circ,7.5^\circ,11.25^\circ,15^\circ\}.
\end{aligned}
\label{eq:planning_problem}
\end{equation}

\begin{figure}[t]
\centerline{\includegraphics[width=\linewidth]{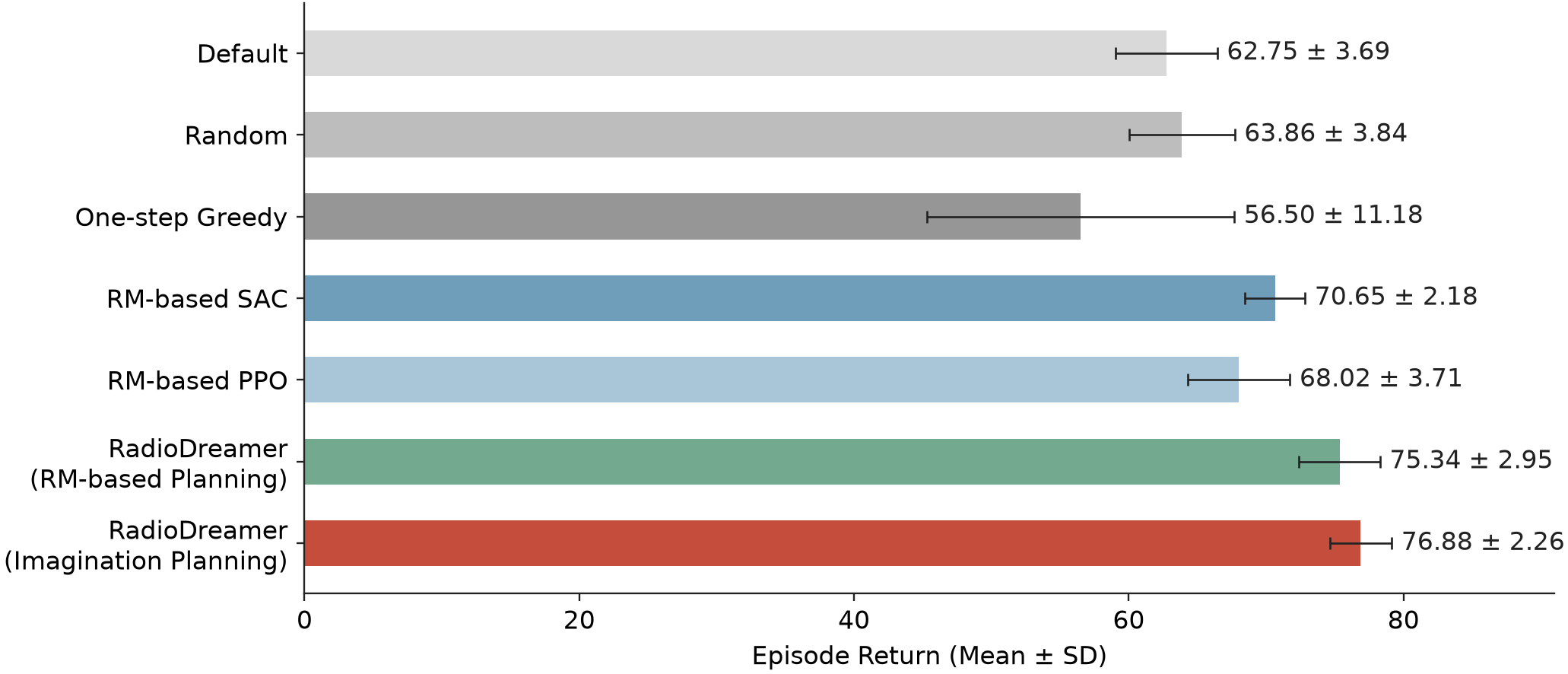}}
\caption{Comparison of Overall Reward Performance.}
\label{fig:planning_return}
\vspace{-0.2cm}
\end{figure}

We compare RadioDreamer with the default BS configuration, random feasible
actions, one-step greedy optimization, RM-based SAC~\cite{haarnoja2018soft},
RM-based PPO~\cite{schulman2017proximal}, and a RadioDreamer variant that plans
over decoded radio maps. As shown in Fig.~\ref{fig:planning_return},
imagination planning achieves the highest episode return of
$76.88\pm2.26$, improving over the default configuration and one-step greedy
optimization by 22.5\% and 36.1\%, respectively. It also outperforms RM-based
SAC by 8.8\% and decoded-radio-map planning by 2.0\%.
Table~\ref{tab:planning_coverage_performance} reports the corresponding
coverage metrics. RadioDreamer achieves an RSRP coverage of $95.05\%$, an SINR
coverage of $71.45\%$, and a 5\%-worst-region coverage of $41.90\%$, ranking
first on all three metrics. Compared with the default configuration, these
values increase by 1.58, 40.01, and 23.55 percentage points, respectively,
showing consistent improvements in both overall and weak-region coverage.
These results demonstrate the benefit of imagination-based planning. Unlike
RM-based SAC and PPO, which directly optimize policies from radio-map
observations, RadioDreamer first learns the action-conditioned evolution of
radio states and then optimizes BS actions through imagined trajectories. Its
advantage over decoded-radio-map planning further suggests that the learned
latent space better captures temporal dependencies and supports more reliable
evaluation of the long-term effects of BS configurations.

\begin{table}[t]
    \centering
    \caption{ Performance of different methods on individual metrics.
    }
    \label{tab:planning_coverage_performance}
    \setlength{\tabcolsep}{3.5pt}
    \renewcommand{\arraystretch}{1.10}

    \resizebox{\columnwidth}{!}{%
    \begin{tabular}{@{}lccc@{}}
        \toprule
        \textbf{Method}
        & \textbf{RSRP Coverage $\uparrow$}
        & \textbf{SINR Coverage $\uparrow$}
        & \textbf{Cell-Edge Coverage $\uparrow$} \\
        \midrule

        Default
        & $93.47 \pm 0.33\%$
        & $31.44 \pm 11.25\%$
        & $18.35 \pm 4.69\%$ \\

        Random
        & $87.92 \pm 2.03\%$
        & $45.63 \pm 9.72\%$
        & $18.92 \pm 4.95\%$ \\

        One-step Greedy
        & $82.19 \pm 6.20\%$
        & $62.59 \pm 10.47\%$
        & $38.45 \pm 5.55\%$ \\

        RM-based SAC
        & $92.99 \pm 0.51\%$
        & $70.24 \pm 6.99\%$
        & $31.89 \pm 5.72\%$ \\

        RM-based PPO
        & $94.11 \pm 0.18\%$
        & $45.39 \pm 11.34\%$
        & $19.01 \pm 5.90\%$ \\

        RadioDreamer (RM-based Planning)
        & $94.97 \pm 0.82\%$
        & $67.13 \pm 7.25\%$
        & $35.15 \pm 7.01\%$ \\

        \textbf{RadioDreamer (Imagination Planning)}
        & $\mathbf{95.05 \pm 0.48\%}$
        & $\mathbf{71.45 \pm 6.96\%}$
        & $\mathbf{41.90 \pm 6.24\%}$ \\

        \bottomrule
    \end{tabular}%
    }
\end{table}

\subsection{Ablation and model complexity}

We conduct ablation experiments on the Street Canyon scene under T1 by
removing Region-Adaptive Supervision, temporal causal attention, BS-level
attention, spatial attention, and historical motion information. As shown in
Table~\ref{tab:ablation}, temporal causal attention contributes the most. At
$H=16$, removing it increases the RSRP/SINR F-MAE from $0.893/1.066$ to
$1.313/1.656$, and raises the D-MAE from $2.584$ to $5.118$. This confirms the
importance of modeling causal dependencies among historical radio states, BS
actions, and future evolution.

Removing Region-Adaptive Supervision increases the RSRP/SINR F-MAE at $H=16$
to $1.131/1.349$ and nearly doubles the D-MAE to $5.104$, showing its benefit
for long-horizon and dynamic-region forecasting. Removing BS-level or spatial
attention also degrades performance, while excluding historical motion
information increases the D-MAE to $3.197$. These results validate the roles
of inter-BS interactions, structured spatial propagation, and mobility cues.

\begin{figure}[t]
\centerline{\includegraphics[width=.8\linewidth]{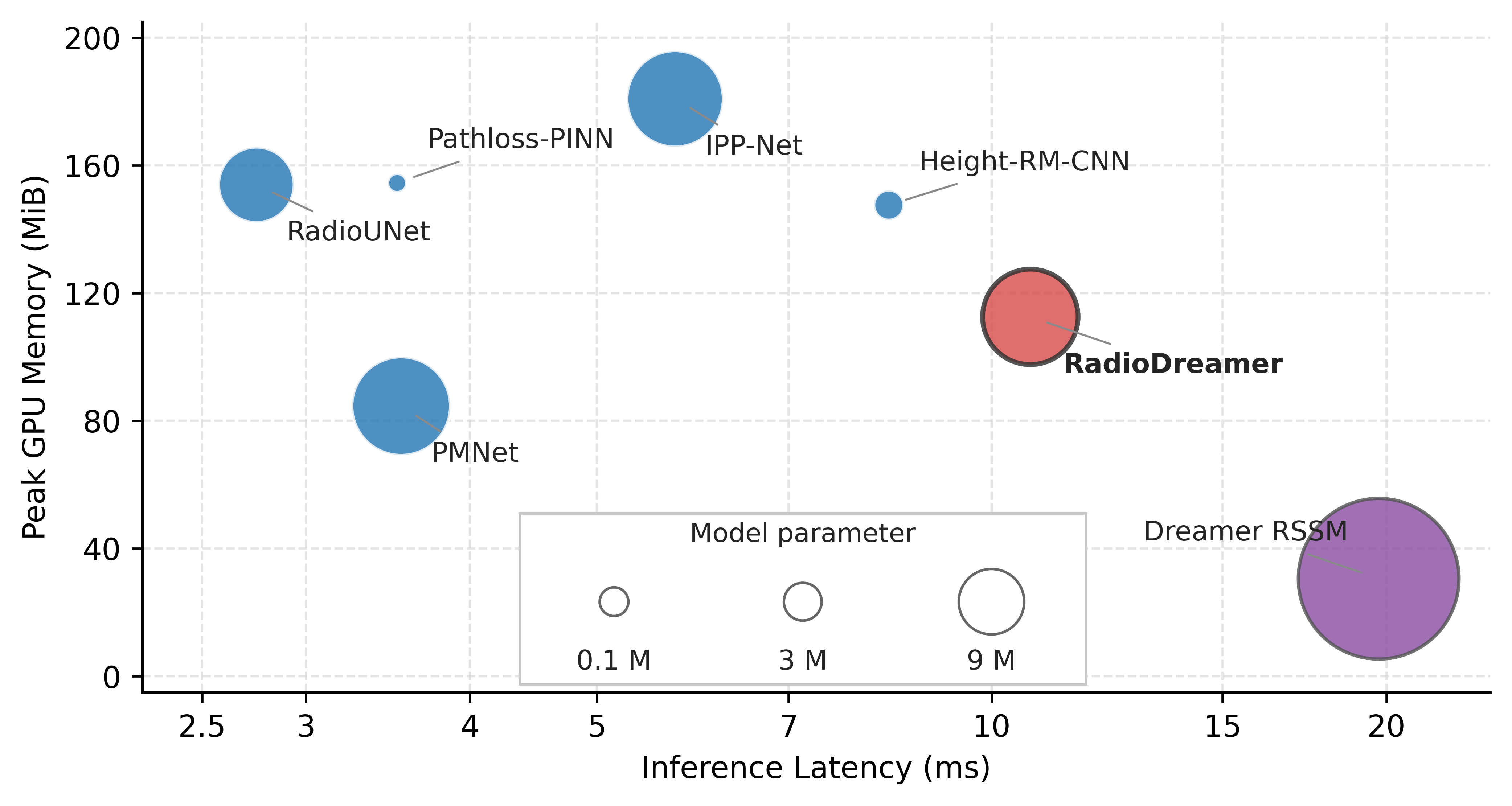}}
\caption{Model Complexity Comparison}
\label{fig:model_complexity}
\vspace{-0.2cm}
\end{figure}

Fig.~\ref{fig:model_complexity} compares inference latency, peak GPU memory,
and trainable parameters. RadioDreamer requires approximately $10.5$~ms per
inference and $112$~MiB of peak GPU memory. Its latency is about half that of
Dreamer-RSSM, and its memory consumption is lower than several static
radio-map models. Overall, RadioDreamer achieves a favorable balance between forecasting accuracy and computational cost.

\begin{table}[t]
    \centering
    \caption{Ablation study iN Street Canyon. 'w/o' means without.}
    \label{tab:ablation}
    \resizebox{\linewidth}{!}{
        \begin{tabular}{lcccc}
            \toprule
            \textbf{Model}
            & \textbf{RSRP / SINR}
            & \textbf{RSRP / SINR}
            & \textbf{RSRP / SINR}
            & \textbf{RSRP} \\
            & F-MAE(H=1)
            & F-MAE(H=8)
            & F-MAE (H=16)
            & D-MAE (H=16) \\
            \midrule
            \textbf{RadioDreamer}
            & \textbf{0.905 / 1.081}
            & \textbf{0.894 / 1.068}
            & \textbf{0.893 / 1.066}
            & \textbf{2.584} \\

            w/o Dynamic Super.
            & 1.017 / 1.190
            & 1.126 / 1.342
            & 1.131 / 1.349
            & 5.104 \\

            w/o Temporal Atten.
            & 1.275 / 1.603
            & 1.304 / 1.649
            & 1.313 / 1.656
            & 5.118 \\
            
            w/o BS Atten.
            & 0.963 / 1.144
            & 0.966 / 1.150
            & 0.967 / 1.150
            & 2.921 \\
            
            w/o Spatial Atten.
            & 0.963 / 1.130
            & 0.968 / 1.141
            & 0.968 / 1.140
            & 2.976 \\
            
            w/o Motion Info.
            & 0.986 / 1.165
            & 0.991 / 1.176
            & 0.990 / 1.174
            & 3.197 \\
            \bottomrule
        \end{tabular}
    }
\end{table}

\section{Conclusion}
This paper presented RadioDreamer, a world model for learning the
action-conditioned dynamics between radio environments and BS configurations
in the latent space. 
By combining structured attention mechanisms with region-adaptive supervision,
RadioDreamer learns an informative latent representation of radio-environment
evolution and supports multi-step imagination for network optimization.
Experiments across three urban scenes show that RadioDreamer reduces radio-map
forecasting error by 25.2\% on average over the strongest baselines,
demonstrating the effectiveness of world models for dynamic radio-map
forecasting. RadioDreamer further supports network optimization through latent
imagination, achieving an 8.8\% higher episode return than planning over decoded
radio maps and validating the advantage of optimizing long-term radio-state
dependencies directly in the latent space. This work represents an initial
attempt to introduce world models into wireless communications. Future work
will incorporate real-world measurements, unified training across multiple
urban scenes, and more comprehensive network planning and optimization tasks.

\clearpage
\bibliographystyle{IEEEtran}
\bibliography{referInfo}

\end{document}